\documentstyle[mprocl]{article}

\def\be{\begin{equation}}
\def\ee{\end{equation}}
\def\bea{\begin{eqnarray}}
\def\eea{\end{eqnarray}}
\def\be{\begin{equation}}
\def\en{\end{equation}}
\def\bear{\begin{eqnarray}}
\def\enar{\end{eqnarray}}
\def\beas{\begin{eqnarray*}}
\def\enas{\end{eqnarray*}}
\def\bera{ \setcounter{enumi}{\value{equation}} 
           \addtocounter{enumi}{1}
           \setcounter{equation}{0}
           \renewcommand{\theequation}{\theenumi\alph{equation}}
           \begin{eqnarray} }
\def\enra{ \end{eqnarray}
           \setcounter{equation}{\value{enumi}}
           \renewcommand{\theequation}{\arabic{equation}}  }
\def\re{\Re e}

\def\ep{\epsilon}

\def\mn{{\mu\nu}}
\def\CA{{\cal A}}

\def\CC{{\cal C}}

\def\den{e}
\def\dtri{\tilde{E}}

\def\been{\begin{enumerate}}
\def\enen{\end{enumerate}}
\def\beit{\begin{itemize}}
\def\enit{\end{itemize}}
\def\bece{\begin{center}}
\def\ence{\end{center}}
\def\bert{\begin{flushright}}
\def\enrt{\end{flushright}}
\begin{document}
\renewcommand{\thefootnote}{\fnsymbol{footnote}}
\setcounter{footnote}{1}

\title{Lorentzian dynamics in the Ashtekar gravity
%\footnote{The Proceedings of the 8th Marcel Grossmann Meeting,
%Jerusalem, June 1997 (World Scientific Press); gr-qc/9710074 }
}

\author{Hisa-aki Shinkai$~^{\dag}$
%\footnote{Electronic address: shinkai@wurel.wustl.edu} 
and Gen Yoneda$~^{\ddag}$ }

\address{
$~^{\dag}$ {\em Dept. of Physics, Washington University, St. Louis, 
MO 63130-4899, USA} \\
$~^{\ddag}$ {\em Dept. of Mathematics, Waseda University, Okubo 3-4-1, 
Shinjuku, Tokyo 169, Japan}
}

%-----------------------------------------------  abstract  >>>
\maketitle\abstracts{
We examine the advantages of the $SO(3)$-ADM (Ashtekar) 
formulation of general relativity, from the point of 
following the dynamics of the Lorentzian spacetime
in direction of applying this into numerical relativity. 
We describe our strategy how to treat new constraints and
reality conditions, together with a proposal of new variables. 
We show an example of passing a degenerate point in flat
spacetime numerically by posing `reality recovering' conditions
on spacetime. We also discuss some 
available advantages in numerical relativity. 
}

%-----------------------------------------------  abstract  <<<

%\baselineskip .24in
%\baselineskip .22in

%\renewcommand{\thefootnote}{\fnsymbol{footnote}}
%\renewcommand{\thefootnote}{\arabic{footnote}}
\setcounter{footnote}{0}
%-----------------------------------------------------------------------
%-----------------------------------------------  paper  >>>
%***********************************************************************
%23456789012345678901234567890123456789012345678901234567890123456789012

%-----------------------------------------------------  introduction
\section{Introduction}

A decade has passed since the proposal of the new formulation of general
relativity by Ashtekar \cite{Ashtekar}. By using the special pair of
variables, the framework has many advantages. 
That is, the constraint equations which appear in the theory become 
low-order polynomials
and do not contain the inverses of the variables, which enables us to 
treat the degenerate points. The theory also has the
correct form for gauge 
theoretical features, and suggests possibilities for treating a quantum 
description of gravity nonperturbatively.

We examine these advantages from the point of 
numerical relativity.  
Salisbury {\it et. al.} \cite{Salisbury} showed a set of equations
using Capovilla-Dell-Jacobson (CDJ) \cite{CDJ} version of connection
formulation for vacuum plane symmetric spacetime for numerical 
demonstration. However, their treatment of reality conditions and
slicing condition is not general enough, and not clear what advantages
can we get in numerical treatment. 
%In this note, 
Here, we introduce our strategy how to treat new constraints and
reality conditions \cite{ys-con} and our study of dynamical treatment of a 
degenerate point \cite{ysn-deg}.  
%We summarize briefly 
%advantages and disadvantages
%of this formulation in the applications to numerical relativity. 

\section{Ashtekar formulation}

The key feature of Ashtekar's formulation is the introduction of a self-dual 
connection as one of the basic dynamical variables.
Let us write\footnote{
We use $\mu,\nu=0,\cdots,3$ and 
$i,j=1,\cdots,3$ as spacetime indexes, while 
$I,J=(0),\cdots,(3)$ and 
$a,b=(1),\cdots,(3)$ are $SO(1,3)$, $SO(3)$ indexes respectively. 
} the metric $g_\mn$ using the tetrad, $e^I_\mu$, and 
define its inverse, $E^\mu_I$, by 
$g_{\mu\nu}=e^I_\mu e^J_\nu \eta_{IJ}$ and 
$E^\mu_I:=e^J_\nu g^{\mu\nu}\eta_{IJ}$.
We define a SO(3,C) self-dual connection
$\CA^a_{\mu} 
:= \omega^{0a}_\mu- ({i / 2}) \ep^a_{~bc}\omega^{bc}_\mu,  
%\label{w2A}
$ %\en
where $\omega^{IJ}_{\mu}$ is a spin connection 1-form (Ricci 
connection), $\omega^{IJ}_{\mu}:=E^{I\nu} \nabla_\mu e^J_\nu.$
%Note that the extrinsic curvature, 
%$K_{ij}=-(\delta_i^{~l}+n_in^l)\nabla_ln_j$
%in the ADM formalism, where $\nabla$ is a covariant 
%derivative on $\Sigma$, 
%satisfies the relation $-K_{ij}E^{ja}=\omega^{0a}_{i}$, 
%when the gauge condition $E^0_a=0$ is fixed.
%So 
%$\CA^a_{i}$ is also expressed by
%\be \CA^a_i = -K_{ij}E^{ja}
%-\dsp{i \over 2} \ep^a_{~bc}\omega^{bc}_i. \label{def-ash}
%\en
The lapse function, $N$, and shift vector, $N^i$,
are expressed as $E^\mu_0=({1 / N}, -{N^i / N}$). 

Ashtekar  treated the set  ($\CA^a_{i}$, $\dtri^i_{a}$) 
as basic dynamical variables, where 
$\dtri^i_{a}$ is an inverse of the densitized triad 
defined by $\dtri^i_{a}:=\den E^i_{a}$, and
where $\den:=\det e^a_i$ is a density.
%This pair forms the canonical set
%\bera
%\{ \dtri^i_{a}(x), \dtri^j_{b}(y) \}
%&=& 0, \label{poisson1}\\
%\{ \CA^a_{~i}(x), \dtri^j_{b}(y) \}
%&=& i \delta^j_{~i} \delta^a_{~b} \delta(x-y), \label{poisson2}\\
%\{ \CA^a_{~i}(x),  \CA^b_{~j}(y) \}
%&=&0.\label{poisson3}
%\enra
Then the full set of equations are dynamical equations for $\CA^a_{i}$ and
$\dtri^i_{a}$, together with Hamiltonian constraint ${\cal C}_H$, momentum
constraint  ${\cal C}_M$ and gauge (Gauss) constraint ${\cal C}_G$.
Details are in %Ashtekar's original references 
\cite{Ashtekar} or 
in \cite{ysn-deg} with cosmological constant.

%---------------------------------------------- Part I.
%---------------------------------------------- Part I.
%---------------------------------------------- Part I.
\section{Reality conditions and additional constraint equation}
If we compare this formulation with conventional Arnowitt-Deser-Misner (ADM)
3+1 formulation, the bottleneck is 
additional constraint ${\cal C}_G$ and the reality conditions. 
The reality conditions are, so far, posed on the metric or the triad. 
We clarified these differences and showed that the triad reality condition
restricts three more freedom than the metric reality condition,
which fixes the real part of the gauge function $\CA^a_{~0}$ 
(we named this function 
{\it triad lapse}) \cite{ys-con}. 

Since we are only interested in the dynamics of real Lorentzian spacetime, 
we have to impose metric reality condition.  From the fact that the reality
of the spacetime is conserved if we solve reality conditions 
initially\cite{Ashtekar89a}, so we propose to prepare ADM initial data for
evolution in Ashtekar's variables by transforming variables and introducing
internal variables as they satisfy ${\cal C}_G$. 

CDJ  solved ${\cal C}_H$ and ${\cal C}_M$
by introducing new variables, which 
corresponds to the Weyl curvature $\Psi_i$. 
In contrast to CDJ, we make an alternative treatment of
the gauge constraint ${\cal C}_G$ and the secondary metric 
reality condition. We summarize these properties in Table 1. 
Details are in \cite{ys-con}.

%##############################  Table.\ref{t1}  >>>>>.
\begin{table}[h]
%\vspace*{1.0cm}
\begin{center}
\begin{tabular}{ c||c|c|c|c|c}
%-----------  0 dan
\hline \hline
&&\multicolumn{2}{c|}{}&&\\
%-----------  1 dan
%~
& {\bf ADM} &\multicolumn{2}{c|}{\bf Ashtekar} & {\bf CDJ} \cite{CDJ}
& {\bf YS} \cite{ys-con}
 \\ [1em]
%-----------  2 dan line
\hline \hline
%&&\multicolumn{2}{c|}{}&&\\[-.1em]
%-----------  2 dan
\begin{tabular}{l} dynamical  \\ variables \end{tabular} & 
$\gamma_{ij}, K_{ij}$ & 
\multicolumn{2}{c|}{$\CA^a_{~i}, \dtri^i_a$}  & 
$\CA^a_{~i}, \Psi_{ab}$ & $\re [\CA^{(ab)}], \dtri^i_a$
\\
%[1em]
%-----------  3 dan line
\hline
%&&\multicolumn{2}{c|}{}&&\\[-.1em]
%-----------  3 dan
\begin{tabular}{l} gauge fixing  \\ variables \end{tabular} & 
$N, N^i$ & 
\multicolumn{2}{c|}{$N, N^i,\CA^a_{~0} $  } & 
$N, N^i,\CA^a_{~0} $  & $N, N^i,\CA^a_{~0} $ 
\\
%[1em]
%-----------  4 dan line
\hline
%&&\multicolumn{2}{c|}{}&&\\[-.1em]
%-----------  4 dan
constraints & $\CC_H, \CC_M$ &  
\multicolumn{2}{c|}{$\CC_H, \CC_M, \CC_G$}   &
 $\CC_G$  & 
%\begin{tabular}{c}
$\CC_H, \CC_M$
%\\ 
%($\CC_G \rightarrow \CA^{[ab]}$) 
%\end{tabular}
\\
%[1em]
%-----------  5 dan line
\hline
%&&&&&\\[-.1em]
%-----------  5 dan
\begin{tabular}{l} reality  \\ conditions \end{tabular} & (none) & 
\begin{tabular}{c}{\bf metric}  \\ primary \\ 2nd \end{tabular} &
\begin{tabular}{c}{\bf triad}   \\ primary \\ 2nd \end{tabular} &
\begin{tabular}{c}{\bf metric}  \\ primary  \\ 2nd \end{tabular} &
\begin{tabular}{l}{\bf triad} primary  \\ {\bf metric}  2nd \\
%~~~
%$\rightarrow \im[\CA^{(ab)}]$ 
(solved)\end{tabular} \\ 
%&&&&&\\[-.1em]
\hline \hline
\end{tabular}
\end{center}
\caption[t1]{A list of 
alternative approaches for time evolution of the three-hypersurfaces.
}
\label{t1}
\end{table}
%##############################  Table.\ref{t1}  <<<<<.

%---------------------------------------------- Part II ----------------
%---------------------------------------------- Part II ----------------
%---------------------------------------------- Part II ----------------
\section{Trick for passing a degenerate point}
Next, we examine the possibilities of
passing a degenerate point. 
A `degenerate point', we use here, is defined as the point 
in the spacetime where
the density		$\den$ %, where	 $\den^2=\det \dtri^i_a$, 
of 3-space vanishes.
In the Ashtekar formulation,  all the equations do not include any
inverse of $\den$ apparently, so that we expect we can `pass'
such a degenerate point. 

In order to say `pass' degenerate points, we start from requiring
the finiteness of the fundmental variables (and their derivatives), 
$\dtri^i_a, \CA^a_i, N/e, N^i, \CA^a_0$, and the condition that
the calculation must be finished in finite coordinate time.
Although these are natural conditions for pursuiting the evolutions 
of spacetime, we concluded that continuing evolutions including a 
degenerate point
in its foliation of 3-space is generally break one of above conditions.
The difficulties are that the term $\omega^{bc}_i$
in $\CA^a_i$ diverges generally and a requirement 
of finite coordinate time
fails when we pass a degenerate point.
%This is simply come from the fact that a requirement of finite coordinate time
%fails when we pass a degenerate point.  
This means generally we face a trouble when we pass a degenerate point directly
in Lorentzian spacetime  
even if we use Ashtekar's variables. 

However, since the variables are originally defined as complex numbers, 
if we are allowed to break the reality condition locally in the neibour of
a degenerate point, which we also assume its degeneracy 
exists only on the real section of spacetime, then we can `pass' a degenerate 
point by such a `deformed slice approach'. Note that, in our proposal,
 the foliation maintains 
$3 + 1$ dimensions 
$\mbox{\boldmath $R$}^3 \times \mbox{\boldmath $R$}$ in 
$\mbox{\boldmath $C$}^4$.

In order to recover a real metric spacetime again later, we have to impose
`reality recovering condition' on the foliation, which requires us to 
determine shooting parameters in complex part of gauge variables.  We 
showed this technique actually works, by demonstrating a numerical evolution
for an analytic solution of degenerate point in flat spacetime\cite{ysn-deg}. 
We see that the time evolution does work properly 
in the sense that
the real part of evolution recovers the analytic evolutions 
and 
the imaginary part of metric vanishes asymptotically.

%---------------------------------------------- subsection 3.3
%---------------------------------------------- subsection 3.3
%---------------------------------------------- subsection 3.3
\section{Discussion}
In summary, when we apply Ashtekar's connection approach to Lorentzian
dynamics, especially in numerical treatment, 
expected difficulties such as treatment of reality conditions and 
additional constraints are conquered by choosing alternative variables
and by preparing ADM initial data. One expected advantage of tractability 
of a degenerate metric requires us to break reality condition locally, but
we found a trick to do so. 

In the last, we comment on another expected advantages in the 
applications for numerical relativity: new available slicing conditions. 
Since we have the additional gauge freedom of 
`triad lapse' $\CA^a_0$, there are wide varieties in choosing a
slicing condition, if we define it using connection variables. 
(Note that just rewriting a slicing
condition defined in ADM in terms of connection variables has no
practical advantages. )
For example, Ashtekar variable has close relation 
with connection or curvature quantities such as 
Newman-Penrose variables, we expect that we can control 
curvature or shear locally more directly than ADM variables.
A detailed discussion together with numerical demonstrations
will be reported elsewhere. % \cite{sy-num}.
We expect that this connection approach to numerical relativity will 
enable us to study also the dynamics of the signature changing process, 
topology changing
process and causal structure in a complex manifold.

%For example, by keeping 
%$\tilde{\CA}:=\CA^a_i \dtri^i_a$ or 
%${\CA}:=\CA^a_i E^i_a$ constant, we can define a similar lapse condition 
%with constant-mean-curvature slicings. 

%\vspace{0.4cm}

%\section*{Acknowledgments}
A part of this work has done with Akika Nakamichi.  
We thank Keiichi Maeda for suggesting us this topic. 
This work was supported in part by NSF PHY 96-00507, PHY 96-00049,
and
by NASA ESS/HPCC CAN NCCS5-153.

%-------------------------------------------------- references -------

\end{document}